----------
X-Sun-Data-Type: default
X-Sun-Data-Name: DiffJetBB.tex
X-Sun-Content-Lines: 1785

\documentstyle[preprint,tighten,floats,prd,aps]{revtex}
\def \DESepsf(#1 width #2){\bf #1  here: just uncomment the macro.}
\begin{document}
\preprint{\vbox{\hbox{OITS 536}\hbox{AZPH-TH/94-02}}  }
\draft
\title{Diffractive jet production in a simple model\\
with applications to HERA}
\author{Arjun Berera}
\address{Department of Physics,
University of Arizona,
Tuscon, AZ  85721}
\author{Davison E.\ Soper}
\address{Institute of Theoretical Science,
University of Oregon,
Eugene, OR  97403}
\date{7 March 1994}
\maketitle
\begin{abstract}
In diffractive jet production, two high energy hadrons $A$ and
$B$ collide and produce high transverse momentum jets, while hadron
$A$ is diffractively scattered. Ingelman and Schlein predicted this
phenomenon.  In their model, part of the longitudinal momentum
transferred from hadron $A$ is delivered to the jet system, part is
lost.  Lossless diffractive jet production, in which all of this
longitudinal momentum is delivered to the jet system, has been
discussed by Collins, Frankfurt, and Strikman. We study the
structure of lossless diffractive jet production in a simple model.
The model suggests that the phenomenon can be probed experimentally
at HERA, with $A$ being a proton and  $B$ being a bremsstrahlung
photon with virtuality $Q^2$. Lossless events should be present for
small $Q^2$, but not for $Q^2$ larger than $1/R_{\rm P}^2$, where
$R_{\rm P}$ is a characteristic size of the pomeron.
\end{abstract}
\pacs{}
%
%
\section{Introduction}

In high energy $p\bar p$ collisions at hadron colliders, one
sometimes produces high transverse momentum jets in the final state.
Here high transverse momentum means, say, transverse momentum greater
than 20 GeV. These jets are interpreted as the decay products of
quarks and gluons (partons) produced by the hard scattering of two
partons from the incoming hadrons.  Cross sections for
\begin{equation}
A+B \to {\rm jets} + X
\end{equation}
calculated in QCD according to this picture match the observations
well \cite{eks,cdfjets}. Normally the two hadrons $A$ and $B$ are
broken up in this process.  It is clear that this should
be so, since a $20 {\rm\ GeV}$ momentum transfer is quite sufficient
to disrupt a hadron.  Nevertheless, in 1985 Ingelmann and
Schlein\cite{is} predicted that events of the type
\begin{equation}
A+B \to A + {\rm jets} + X,
\label{dhs}
\end{equation}
where hadron $A$ is diffractively scattered, should occur with a small
but not tiny probability.  Here by ``diffractively scattered,'' we mean
that $A$ emerges with a fraction $(1 - z) > 0.9$ of its original
longitudinal momentum and with a small transverse momentum
$|{\bf P}_{\!A}^{\prime}| \lesssim 1 {\rm\ GeV}$.  The transverse
momentum transfer can also be characterized using the invariant
momentum transfer $t$ from the hadron: $t = (P_{\!A} -
P_{\!A}^\prime)^2 = - ({\bf P}_{\! A}^{\prime\, 2} + z^2 M_A^2) /
(1-z) \approx - {\bf P}_{\! A}^{\prime\, 2}$.

The picture for such diffractive hard scattering proposed by Ingelman
and Schlein is that hadron $A$ exchanges a pomeron with the rest of
the system, where ``pomeron'' means whatever is exchanged in elastic
scattering at large $s$, small $t$. Thus the cross section is
proportional to the pomeron coupling to hadron $A$ as measured in
elastic scattering. The pomeron carries transverse momentum $-{\bf
P}_{\! A}^{\prime}$ and a fraction $z$ of the hadron's longitudinal
momentum.  Here we do not need to know what a pomeron is, only that
its momentum is carried by quarks and gluons. One of these collides
with a parton from hadron $B$ to produce the jets.  Let the parton
that participates in the hard scattering carry a fraction $x$ of the
longitudinal momentum of the incoming hadron $A$, and thus a fraction
$x/z$ of the longitudinal momentum transferred by the pomeron.  Then
the cross section in this model is proportional to a function
$f_{a/P}(x/z,t;\mu)$, where $f_{a/P}(\xi,t;\mu)\, d\xi$ is interpreted
as the probability to find a parton of kind $a$ in a pomeron, where
the parton carries a fraction $\xi$ of the pomeron's longitudinal
momentum. In principle, $f_{a/P}$ can depend on the invariant momentum
transfer $t$, and it should also depend on a scale parameter $\mu$
that characterizes the scale of virtualities or transverse momenta in
the hard parton scattering.

Several guesses for the $\xi$ dependence of $f_{a/P}(\xi,t;\mu)$ were
given in the literature \cite{is,guess,bcss}, but ultimately it was left
to experiment to measure these functions.  An analysis of the
theoretical expectations and of the approximate factorization
hypothesis that is inherent in the model of Ingelman and Schlein was
given in \cite{bcss}.

The reaction (\ref{dhs}) anticipated by Ingelman and Schlein has been
seen at the CERN collider by the UA8 experiment \cite{UA8}. However,
the experiment suggests a feature not anticipated in
\cite{is,guess,bcss}. It was expected that the functions
$f_{a/P}(x/z)$ would have support only for $x<z$.  That is, some of
the momentum fraction $x$ transferred from hadron $A$ would be lost,
appearing in low $P_T$ particles instead of the jets.  Instead, the
experiment suggests that a fraction of the events are lossless in the
sense that $x=z$.  It is as if the formula for the cross section
contained a term proportional to $\delta(1-x/z)$.

It will be important to confirm the existence of a term that is
proportional to $\delta(1-x/z)$ or, more generally, that is singular
as $x/z \to 1$. The UA8 experiment demonstrates that something is
present in addition to terms that vanish as $x/z \to 1$.  However,
given the experiment's rather limited resolution, it is not clear to
us that the experimental findings could not be fit with a function
that is merely finite as $x/z \to 1$. We shall present evidence in a
forthcoming paper that such terms are to be expected \cite{bs}.

Recently, the Zeus and H1 experiments at HERA have reported
the first evidence for the analogous reaction in deeply inelastic
electron scattering \cite{zeusH1},
\begin{equation}
e + A \to e + A + X.
\label{ddis}
\end{equation}
Here the diffractively scattered proton $A$ has not been seen, but there
is evidence for the ``rapidity gap'' expected in such events.
Information on the $x$ and $z$ dependence of the cross section is not yet
available. We will discuss HERA physics later in this paper,
concentrating on $e + A \to e + A + X$ events in which the virtual
photon has low virtuality, rather than high virtuality.

The purpose of this paper is to consider some of the theoretical
issues in light of the experimental results.  In particular, we
reexamine the factorization hypothesis of \cite{is}, and argue that
this hypothesis is at best approximate. This conclusion is in
agreement with that of Collins, Frankfurt, and Strikman\cite{cfs},
although our analysis of the source of the breakdown of
Ingelman-Schlein factorization will be presented in more depth than
that of Ref.~\cite{cfs}. We frame much of the analysis in terms of a
simple model that allows a detailed examination of the theoretical
issues.

Our analysis suggests three conclusions.  First, the factorization
model with a ``distribution of partons in a pomeron'' may ultimately
not be a fruitful way of understanding the process.  Second, the
lossless events found in hadron-hadron collisions should
be absent in deeply inelastic scattering.

Third, perhaps the most interesting conclusion suggested by the model is
that if one examines the process $e + A \to e + A + X$ as a function
of the virtuality $Q^2$ of the virtual photon emitted by the
electron, then the lossless events should be present for
$Q^2 = 0$ but disappear gradually as $Q^2$ becomes larger than some
value $1/R_{\rm P}^2$.  The length scale $R_{\rm P}$ may be interpreted
in the model as the transverse ``size'' of the pomeron.

Our discussion of the experimental possibilities at HERA bears some
resemblance to that found in a recent paper of Brodsky, Frankfurt,
Gunion, Mueller, and Strikman \cite{Gunion}.  These authors study
$e + A \to e + A + V + X$, where $A$ is a proton and $V$ is a vector
meson.  They look at the region $1\ {\rm GeV}^2\ \ll Q^2 \ll s$. As in
the present paper, large $Q^2$ implies a small size for the wave
function of the quarks produced by the virtual photon. As in the
present paper, these quarks couple to the proton via two gluons

\section{Diffractive jet production}

Let us consider, as a definite example, the inclusive cross section
for the production of two jets in a high energy collisions of two
hadrons, $A$ and $B$. Let the initial hadron $A$ have momentum
\begin{equation}
P_{\!A}^\mu = (P_{\!A}^+,P_{\!A}^-,{\bf P}_{\!A})
=  (P_{\!A}^+, {M^2 \over 2P_{\!A}^+},{\bf 0})\,,
\end{equation}
where we denote $P^\pm = (P^0 \pm P^3)/\sqrt 2$ and where we denote
transverse vectors by boldface letters. Similarly, hadron $B$ enters
the scattering with momentum
\begin{equation}
P_{\!B}^\mu = (P_{\!B}^+,P_{\!B}^-,{\bf P}_{\!B})
=  ({M^2 \over 2P_{\!B}^-},P_{\!B}^-, {\bf 0})\,.
\end{equation}
We specify the two jets by variables $E_T$, $X_A$, and $X_B$, given in
terms of the four momenta $P_{\!1}^\mu$ and $P_{\!2}^\mu$ of jets
$1$ and $2$ by
\begin{eqnarray}
E_T &=& (|{\bf P}_{\!1}| + |{\bf P}_{\!2}|)\,,
\nonumber\\
X_A &=& (P_{\!1}^+ + P_{\!2}^+)/P_{\!A}^+\,,
\nonumber\\
X_B &=& (P_{\!1}^- + P_{\!2}^-)/P_{\!B}^-\,.
\end{eqnarray}

The production of high transverse momentum jets is a hard process,
to which QCD factorization applies.  Thus the total cross section
for inclusive two jet production can be written as
\begin{eqnarray}
\lefteqn{{d \sigma(A + B\to {\rm jets} +X)\over d E_T\, dX_A\, dX_B}
\sim }
\nonumber\\ &&
\sum_{a,b}
\int dx_a\, f_{a/A}(x_a;\mu)
\int dx_b\, f_{b/B}(x_b;\mu)\
{d \hat\sigma(a + b\to {\rm jets} +X) \over d E_T\, dX_A\, dX_B}\,.
\label{usualfact}
\end{eqnarray}
Corrections to this asymptotic equality are suppressed by a power of
$m/E_T$, where $m$ represents a typical hadronic momentum scale,
say 300 MeV. Here $d \hat \sigma$ is the parton-level cross section,
computed in perturbation theory with a certain prescription for
removing infrared divergences. The functions $f_{a/A}(x_a;\mu)$ are the
parton distribution functions.  Notice that the Born-level cross
section $d\hat \sigma$ is proportional to $\delta(x_a -
X_A)\,\delta(x_b - X_B)$ since, at the Born level, all of the momentum
carried by the colliding partons is transferred to the observed jets.
In higher orders of perturbation theory, one will have $X_A \le x_a$
and $X_B \le x_b$.

The topic of this paper is inclusive jet production in
which we make a very restrictive demand on the final state.
We demand that hadron $A$ appear in the final state with momentum
$P_{\! A}^\prime$ given by
\begin{equation}
P_{\!A}^{\prime\mu}
=  ((1-z)P_{\! A}^+, {M^2 +{\bf P}_{\! A}^{\prime\, 2} \over
2(1-z)P_{\! A}^+},{\bf P}_{\! A}^{\prime})\,,
\end{equation}
The invariant momentum transfer to hadron $A$ is
\begin{equation}
t = (P_{\!A}-P_{\!A}^\prime)^2
= - {1 \over 1-z} \left\{ {\bf P}_{\! A}^{\prime\, 2} + z^2 M^2
\right\}\,.
\end{equation}
For diffractive scattering, we require that $t$ be small compared to
$s = (P_{\!A}+P_{\!B})^2$ and that $z$ be small, say $z < 0.1$.
Since $X_A \le x_a \le z$, one must look for diffractive hard scattering
in the region $X_A < 0.1$. We thus consider the cross section
\begin{equation}
{d \sigma^{\rm diff}(A + B\to A + {\rm jets} +X)\over
d E_T\, dX_A\, dX_B\, dz\, dt}
\end{equation}
in the region described.

\section{Model of Ingelman and Schlein}

We begin our discussion of diffractive jet production by reviewing
the model of Ingelman and Schlein \cite{is}. We follow much of the
notation of Ref.~\cite{bcss}, but break the presentation into two
stages.  In the first stage, we hypothesize that the cross section
for diffractive jet production can be written in terms of a
diffractive parton distribution:
\begin{eqnarray}
\lefteqn{
{d \sigma^{\rm diff}(A + B\to A +{\rm jets} +X)\over
d E_T\, dX_A\, dX_B\, dz\, dt}
\sim }
\nonumber\\ &&
\sum_{a,b}
\int dx_a\, {d\, f^{\rm diff}_{a/A}(x_a,\mu)\over dz\, dt}
\int dx_b\, f_{b/B}(x_b;\mu)\
{d \hat\sigma(a + b\to {\rm jets} +X) \over d E_T\, dX_A\, dX_B}\,.
\label{factor}
\end{eqnarray}
Here
\begin{equation}
{d\, f^{\rm diff}_{a/A}(x_a,\mu)\over dz\, dt}\ dx_a
\end{equation}
represents the probability to find in hadron $A$ a parton of type $a$
carrying momentum fraction $x_a$, while leaving hadron $A$ intact
except for the momentum transfer $(z,t)$. In the second stage, we
hypothesize that ${d\, f^{\rm diff}_{a/A}(x_a,\mu)/ dz\, dt}$ has a
particular form:
\begin{equation}
{d\,f^{\rm diff}_{a/A}(x_a,\mu)\over
dz\, dt}
={1 \over 16\pi}\, |\beta_{A}(t)|^2 z^{-2\alpha(t)}\,
f_{a/P}(x_a/z, t,\mu)\,.
\label{pomdist}
\end{equation}
Here $\beta(t)$ is the pomeron coupling to hadron A and $\alpha(t)$ is
the pomeron trajectory.  The function $f_{a/P}(x_a/z, t,\mu)$ is then
the ``distribution of partons in the pomeron.''

In writing Eq.~(\ref{pomdist}), one thinks of the pomeron as a
continuation in the angular momentum plane of a set of hadron
states.  Since hadrons contain partons, the pomeron should also. Thus
one has in Eq.~(\ref{pomdist}) the standard factors describing the
coupling of the pomeron to hadron A, together with a distribution of
partons in the pomeron \cite{is,bcss}. Eq.~(\ref{factor}), on the
other hand, is more basic.  It  says only that factorization still
applies when hadron $A$ is diffractively scattered.  It is
Eq.~(\ref{factor}) that will be of concern in this paper. We shall
argue that, in fact, factorization does not generally apply when
hadron $A$ is diffractively scattered.

\section{Factorizing and non-factorizing graphs}

In this section, we discuss the Ingelman-Schlein model in terms of the
contributing Feynman graphs. We first give a graph-based argument that
the Ingelman-Schlein picture is plausible.  Fig.~\ref{diffjetfig} shows
a typical graph that contributes to diffractive jet production. The
shaded circles represent Bethe-Salpeter wave functions for the two
hadrons, $A$ and $B$. In the center of the diagram, there is a tree
level contribution to the hard scattering of two gluons with momenta
approximately collinear to the momenta of their respective parent
hadrons. The hard scattering produces the observed high $E_T$ jets. The
jets carry plus-momentum $X_A P_{\!A}^+$. The gluon that enters the
hard scattering from hadron $A$ carries plus-momentum $x_a P_{\!A}^+$,
which equals $X_A P_{\!A}^+$ for Born-level hard scattering. We may
view this gluon as a normal part of the gluon cloud that accompanies
any hadron.  Since only a small fraction $X_A < 0.1$ of hadron $A$'s
plus-momentum has been lost to the hard scattering, it seems plausible
that the gluon cloud and the constituent quarks can reassemble
themselves into the hadron bound state.  It is only necessary to
dispose of one more gluon, as indicated in the figure.  Then the net
color transfer from hadron $A$ can be zero.  If this gluon carries a
small momentum fraction $x^\prime$ into the final state, then the net
momentum fraction lost from hadron $A$, $z = X_A + x^\prime$, will also
be small ($z< 0.1$, say).  Similarly, the net transverse momentum
$-{\bf P}_{\! A}^{\prime}$ transferred from the hadron can be small
($|{\bf P}_{\! A}^{\prime}| \lesssim 1 {\ \rm GeV} $).

\begin{figure}[htb]
\centerline{ \DESepsf(diffjet.epsf width 10 cm) }
\smallskip
\caption{Diffractive jet production.}
\label{diffjetfig}
\end{figure}

Let us now compare Fig.~\ref{diffjetfig} with a typical graph that
contributes to pomeron exchange, {\it i.e.} hadron-hadron elastic
scattering at large $s$ and small $t$, as depicted in
Fig.~\ref{pomeronfig}. Of course, the pomeron is much more
complicated than this, so the reader is invited to imagine some much
more complicated graphs \cite{bkflpomeron}.  Nevertheless, it seems
an attractive proposition that whatever graphs contribute to the
bottom half of the pomeron in Fig.~\ref{pomeronfig} can contribute
equally to the bottom half of Fig.~\ref{diffjetfig}.

\begin{figure}[htb]
\centerline{ \DESepsf(pomeron.epsf width 10 cm) }
\smallskip
\caption{Typical graph contributing to the pomeron.}
\label{pomeronfig}
\end{figure}

This graphical argument makes the Ingelman-Schlein model seem
plausible.  Since the pomeron-proton coupling is large, the argument
also suggests that the cross section for diffractive jet production
at a given $E_T$ is not an infinitesimal fraction of the total cross
section for jet production with the same jet $E_T$.
In this paper, however, we will focus on two other features of
Fig.~\ref{diffjetfig}, factorization and longitudinal momentum loss.

First, we note that the top half of the graph is connected to the
bottom half only at the hard scattering. The absolute square of the
graph breaks up into a convolution of a factor that contributes to the
distribution function of gluons in hadron $B$, times a contribution to
the hard scattering cross section, times a contribution to the
distribution function of gluons in hadron $A$, with the additional
requirement that hadron $A$ is diffractively scattered. This is just
the structure of the factorized cross section formula (\ref{factor}).
Thus the usual factorization program (as described in
\cite{factorization}) is relatively straightforward for this graph
(although there are some subtleties related to the gluon
polarizations).

Second, we note that, in order to get the color right, it was necessary
to emit a gluon (or more than one gluon) into the final state.
Inevitably, then, some of the net plus-momentum $z P^+$ transferred by
the pomeron is not contributed to the hard scattering, but is lost
into the final state in the form of low transverse momentum particles.
Thus the contribution to the cross section from graphs like
Fig.~\ref{diffjetfig} will be nonzero only for $X_A/z < 1$.

Consider, however, the graph of Fig.~\ref{diffjetmodfig}.  Here the
gluon that had to be emitted in order to make the color right is
absorbed by a parton from hadron $B$, which is moving in the opposite
direction.  As we will see in the following sections, the plus-momentum
transferred by the extra gluon in this process is negligible compared
to $zP^+$.  Thus the contributions from graphs like
Fig.~\ref{diffjetmodfig} will be effectively proportional to
$\delta(1 - X_A/z)$.  Furthermore, graphs like this break the
Ingelman-Schlein factorization represented by Eq.~(\ref{factor}).  As
argued in \cite{factorization}, such factorization breaking effects
cancel when one calculates an inclusive cross section to produce jets
with no restriction on the final state.  However, here there is an
important restriction on the final state: we allow only final
states consisting of an elastically scattered proton.  Thus
factorization breaking effects may survive. As noted in \cite{cfs},
a signature of these effects is the appearance of $\delta(1 - X_A/z)$
terms in the cross section.

\begin{figure}[htb]
\centerline{ \DESepsf(diffjetmod.epsf width 10 cm) }
\smallskip
\caption{Diffractive jet production with color transferred to
spectator parton.}
\label{diffjetmodfig}
\end{figure}

In the following sections, we explain a simple model in which this
effect can be examined in some detail.  Within this model, we find
that the factorization breaking effect has some simple features.  We
abstract the simple features from the model and propose that one can
test for these features in electron-proton scattering at HERA.

\section{A simple model}

We will examine diffractive jet production within the context
of scalar-quark QCD.  In this model, there is an SU(3) gauge field
$A^\mu_a(x)$ as in normal QCD. There is also a color triplet quark
field $q_i(x)$ with quark mass m, but we take $q_i(x)$ to be a scalar
field instead of a Dirac field. Since the theory includes a scalar
quark field, a 4-quark coupling is necessary, but we set the
renormalized coupling constant $g_4$ to a negligibly small value. The
lagrangian is
\begin{eqnarray}
{\cal L} &=&  (D_{\mu}q)^\dagger ( D^{\mu}q) -m^2 q^\dagger q
-{1\over 4} G^{a}_{\mu \nu }G_{a}^{\mu \nu }
-{1\over 2\xi}(\partial_\mu A_a^\mu)(\partial_\nu A_a^\nu)
\nonumber\\
&&+ \ \hbox{\rm Faddeev-Popov terms}
-{g_4\over 4}(q^\dagger q)^2.
\end{eqnarray}
This theory has the same behavior as spinor QCD for collinear and
soft gluon emission from quarks.  Its chief advantage is that it
allows a perturbative model for a quark-antiquark bound state.  We
introduce a scalar, color singlet meson field $\phi(x)$ and couple it
to the quarks using
\begin{equation}
{\cal L}_\phi = G\ \phi(x)\, q^\dagger_i(x) q_i(x).
\end{equation}
We work to lowest nontrivial order in the $\phi q^\dagger q$ coupling
$G$, letting the $\phi q^\dagger q$ vertex play the role that is played
by the (amputated) Bethe-Salpeter wave function of a meson in spinor
QCD. We denote the mass of the meson by $M$ and take $M$ to be smaller
than $2m$, so that the meson cannot decay into a quark and an
antiquark. This model has been used for similar purposes in
Refs.~\cite{lrs} and \cite{cssmodel}.

In the following sections, we examine diffractive jet production
within the context of low order perturbative graphs in this model.
The idea is to understand some of the basic physics by means of
examples.  We will be working with perturbation theory for the cross
section at order $\alpha_s^4 G^6$ and $\alpha_s^5 G^6$. Of course, the
real strong coupling $\alpha_s(\mu)$ is not small except at large
$\mu$, and some parts of our graphs contain soft momentum flows, for
which small $\mu$ would be relevant. We hope (but do not prove)
that the basic lessons we learn here will be valid at order
$\alpha_s^N G^6$ for any $N$ in the model, and at order $\alpha_s^N$
in real QCD. However, such an analysis at arbitrary order $N$
remains a challenge for the future.

\section{Wave functions}

The analysis that follows will make use of the null-plane wave
function of the meson state in our model. For a meson moving in the
minus direction, the wave function $\psi(x,{\bf k})_{ij}$ is the
amplitude to find that the meson with momentum $P^\mu = (M^2/(2P^-),
P^-,{\bf 0})$ consists of a quark and an antiquark of colors $i$ and
anti-$j$ respectively, with the quark having minus-momentum $k^- =
xP^-$ and transverse momentum ${\bf k}$.  The wave function is
measured by operators defined on the null-plane $y^- = 0$.  The
precise definition, following the formalism of
\cite{ks,bks,BrodskyLepage}, is
\begin{equation}
\psi(x,{\bf k})_{ij} =
2x(1-x)P^- \int d^4 y\ e^{ik\cdot y}\, \delta(y^-)\
\langle 0|
q_i(y)\,q^\dagger_j(0)
|P\rangle \,.
\label{psidef}
\end{equation}
Here we have chosen the normalization
\begin{equation}
(2\pi)^{-3}\int_0^1 {dx \over 2x(1-x)}\int d{\bf k} \sum_{ij}
|\psi(x,{\bf k})_{ij}|^2
= P_{\!2}\,,
\label{psinorm}
\end{equation}
where $P_{\!2}$ is the probability, which is of order $G^2$, that the
meson state consists of a $(q,q^\dagger)$ pair.  In terms of the
covariant $\phi\, q\, q^\dagger$ Green function amputated on the
$\phi$-leg, the definition (\ref{psidef}) can be written as
\begin{equation}
\psi(x,{\bf k})_{ij} =
2x(1-x)P^- \int {\ dk^+\over 2 \pi}\, {\cal
G}(k^\alpha,P^\beta)_{ij}\,,
\label{psigreen}
\end{equation}
At lowest order in $\alpha_s$ and $G$ one has,
\begin{equation}
{\cal G}(k^\alpha,P^\beta)_{ij} = iG\,\delta_{ij}\,
{i \over k^2 - m^2 + i\epsilon}\
{i \over (P-k)^2 - m^2 + i\epsilon}
\,.
\end{equation}
By integrating according to Eq.~(\ref{psigreen}), we find
\begin{equation}
\psi(x,{\bf k})_{ij} =
{ G x(1-x)
 \over {\bf k}^2 + m^2 - x(1-x)M^2}
\ \delta_{ij}\,.
\label{psi}
\end{equation}
Notice that $|\psi|^2 \propto 1/{\bf k}^4$ for large ${\bf k}^2$.
This good behavior in the ultraviolet, which arises from the fact that
$G$ has dimensions of mass, is the essential reason for the
usefulness of this model.

The wave function can be used to calculate, to order zero in
$\alpha_s$, the probability for finding a quark in a meson
{\it i.e.} the parton distribution function:
\begin{eqnarray}
f_{q/\phi}(x) &=&
(2\pi)^{-3}{1 \over 2x(1-x)}\int d{\bf k} \sum_{ij}
|\psi(x,{\bf k})_{ij}|^2
\nonumber\\ &=&
{3 G^2\over 16 \pi^2}
{ x(1-x)
\over m^2 - x(1-x)M^2 }\,.
\label{pdf}
\end{eqnarray}

We pause here to consider a more realistic example, chosen in
anticipation of the application to HERA physics in the final section
of this paper. We let the meson be a photon with polarization vector
$\epsilon$ and we let the quarks be genuine spin-$1/2$ quarks with
negligible mass. The perturbative Green function for the photon to
turn virtually into the quark-antiquark system is
\begin{equation}
{\cal G}(k^\alpha; P^\beta)
= ie{\cal Q}\
{ k\cdot\gamma \over k^2 +i\epsilon}\
\epsilon\cdot\gamma\
{ (k-P)\cdot\gamma \over (k-P)^2 +i\epsilon}.
\end{equation}
Here $e$ is the proton charge and ${\cal Q}$ is the quark charge in
units of the proton charge. The Dirac indices are left implicit
and there is an implicit unit matrix in the $3\times3$ color space.
The wave function, defined analogously to Eq.~(\ref{psidef}), is
\begin{equation}
\psi(x,{\bf k}) = {1 \over {\textstyle 2P^-}}
\int
{ dk^+ \over 2\pi}\
\overline U\ \gamma^-{\cal G}(k^+,xP^-,{\bf k}; P^\beta)\ \gamma^- V\,,
\label{photonpsi}
\end{equation}
where $U$ and $V$ are the Dirac spinors for the quark and antiquark,
respectively.

We consider an off-shell photon with $P^2 \equiv - Q^2 <0 $ and zero
transverse momentum, so that
\begin{equation}
P^\mu = (-{Q^2 \over {\textstyle 2P^-}},P^-,{\bf 0}).
\end{equation}
To define the possible polarization states, we use light-like axial
gauge with gauge-fixing vector $v^\mu= (1,0,{\bf 0})$. In this gauge
the photon propagator is $i{\cal N}^{\mu\nu}/(P^2 + i\epsilon)$ where
\begin{equation}
{\cal N}^{\mu\nu}=
- g^{\mu\nu} +{v^\mu P^\nu + P^\mu v^\nu\over P\cdot v}
= \sum_{i = 1}^2 \epsilon(P,i)^\mu \epsilon(P,i)^\nu
-\epsilon(P,L)^\mu \epsilon(P,L)^\nu.
\end{equation}
Here $\epsilon(P,1)^\mu$ and $\epsilon(P,2)^\mu$ are the two
transverse polarization vectors, with $v\cdot \epsilon = P\cdot
\epsilon = 0$ and $\epsilon\cdot \epsilon = - 1$. The longitudinal
polarization vector is
\begin{equation}
\epsilon(P,L)^\mu = {Q \over P\cdot v}\ v^\mu.
\end{equation}
As $Q^2\to 0$, only the transverse polarization contributes.
The longitudinal mode becomes important for $Q^2 \ne 0$.  It does
not propagate, but is responsible for the ``instantaneous
scalar photon exchange'' force \cite{ks,bks}.

We now insert polarization vectors into $\cal G$ and perform the
$k^+$-integration in Eq.~(\ref{photonpsi}).  For transverse
polarization, we find
\begin{equation}
\psi(x,{\bf k}) =
-{e{\cal Q} \over {\textstyle 2P^-}}\
{\overline U\ \left[
x\ \epsilon\cdot\gamma\ k_T\cdot\gamma
- (1-x)\ k_T\cdot\gamma\ \epsilon\cdot\gamma
\right]\gamma^- V
\over {\bf k}^2 + x(1-x)Q^2},
\label{psiT}
\end{equation}
where $k_T^\mu = (0,0,{\bf k})$. This is quite similar to the scalar
wave function, Eq.~(\ref{psi}).  The chief difference is the factor
of ${\bf k}$ in the numerator, which leads to a logarithmic
divergence in the normalization integral for $\psi$. (The spinors $U$
and $V$ depend on ${\bf k}$, but this dependence is eliminated when
the spinors stand next to a $\gamma^-$.) For longitudinal
polarization, we obtain
\begin{equation}
\psi(x,{\bf k}) =
{e{\cal Q} \over {\textstyle P^-}}\
x(1-x)Q\
{\overline U\gamma^- V
\over {\bf k}^2 + x(1-x)Q^2}\,.
\label{psiL}
\end{equation}
Again, $\overline U\gamma^- V$ is independent of ${\bf k}$. This
wave function is small compared to that for transverse polarization
when $Q^2 \ll {\bf k}^2$ but becomes comparable when $Q^2 \sim {\bf
k}^2$.

\section{Lowest order contribution}

We consider the cross section
\begin{equation}
{d \sigma^{\rm diff}(A + B\to A + {\rm jets} +X)
\over
d E_T\, dX_A\, dX_B\, dz\, dt}
\end{equation}
for diffractive jet production in meson-meson collisions within the
model described above. In  Fig.~\ref{goodgraphfig} we show a lowest
order graph for this process. The jets are produced in a quark-gluon
collision.  The active quark in this collision comes from meson B.
The gluon comes from meson A, where it has been emitted from one of
the quark lines.  Another gluon is emitted into the final state and
meson A is reconstituted.  There are eight ways of attaching the
gluons to the scalar quark and antiquark lines; only one of them is
shown. Similarly, there are four lowest order graphs for quark-gluon
scattering. Thus one needs to sum over the 32 Feynman graphs
contributing to the amplitude. There is also an analogous set of
graphs in which the hard collision is between an antiquark and a gluon.

\begin{figure}[htb]
\centerline{\hskip 2cm \DESepsf(goodgraph.epsf width 8 cm) }
\smallskip
\caption{Factorizable graph for diffractive jet production.}
\label{goodgraphfig}
\end{figure}

The contribution to the cross section calculated from graphs like
Fig.~\ref{goodgraphfig} takes the factorized form of
Eq.~(\ref{factor}) in the high energy limit.  This is a simple
application of the general analysis of \cite{factorization}.
We do not give this analysis in any detail, but we do mention
three key points.

First, we discuss the polarization of the gluon carrying momentum
$k_A$ in Fig.~\ref{goodgraphfig}. We consider all graphs to be in
Feynman gauge in this paper, since to use a physical gauge
introduces non-physical singularities into the graphs.  In Feynman
gauge, the relevant part of the graph has the form
\begin{equation}
H^\alpha {-ig_{\alpha\beta} \over k_A^2 + i\epsilon}\ J^\beta\,,
\end{equation}
where $H$ represents the hard scattering part of the graph and $J$
represents the meson wave function part at the bottom of the graph.
Now we can replace $-g^{\alpha\beta}$ by
\begin{equation}
\sum_{\lambda=1}^2
\epsilon^\alpha(k_A^+,{\bf k}_A,\lambda)\
\epsilon^\beta(k_A^+,{\bf k}_A,\lambda)
-{u^\alpha k_A^\beta
+  k_A^\alpha u^\beta
\over u\cdot k_A}
+{ k_A^2 \over (u\cdot k_A)^2}\ u^\alpha u^\beta
\,,
\end{equation}
where $u^\alpha = (0,1,{\bf 0})$ is a lightlike vector in the
minus-direction and $\epsilon^\alpha$ is the polarization vector for a
gluon with momentum $(k_A^+,{\bf k}_A)$ and polarization $\lambda$ in
null-plane gauge, $u\cdot A \equiv A^+= 0$. (See ref.~\cite{ks} for
definitions.) This much is an identity.  Now using the power counting
of ref.~\cite{factorization}, one finds that the terms proportional to
$u^\alpha k_A^\beta$ and $u^\alpha u^\beta$ are negligible compared to
the terms proportional to $\epsilon^\alpha \epsilon^\beta$ in the
leading integration region, in which $k_A$ is approximately collinear
to $P_{\!A}$. Thus these terms can be dropped.  On the other hand,
for each Feynman graph, the term proportional to $k_A^\alpha u^\beta$
is {\em larger} than the terms proportional to $\epsilon^\alpha
\epsilon^\beta$. This is an artifact of Feynman gauge.  When one sums
$H_\alpha k_A^\alpha$ over the graphs contributing to $H$, there is an
almost complete cancellation and the remaining terms are negligibly
small. The result of this is that $- g^{\alpha\beta}$ can be replaced
by $\sum \epsilon^\alpha\,\epsilon^\beta$.  That is, only physical
polarizations contribute.

The second point concerns the gluon carrying momentum $q$ in
Fig.~\ref{goodgraphfig}. The leading integration region for $q$ is
the collinear region, in which $\bf q$ is the same order of magnitude
as the transverse momenta inside the meson: ${\bf q}^2 \sim m^2$. In
an abelian theory, the graphs without this gluon would constitute the
lowest order graphs for diffractive jet production.
Fig.~\ref{goodgraphfig} would represent another contribution, a
contribution that is formally of higher order in $\alpha_s$ but is
still a leading contribution in the sense of {\em not} being
suppressed by a power of $m/E_T$. However, in the non-abelian theory,
Fig.~\ref{goodgraphfig} gives the lowest order contribution because
the emission of a second gluon is necessary to restore the meson to a
color singlet state.

The third point concerns the factor representing meson $B$:
\begin{equation}
iG\delta_{ij}\, {i \over k_B^2 - m^2 + i\epsilon}\ .
\end{equation}
Taking into account that $P_{\!B}^\mu =
(M^2/(2P_{\!B}^-),P_{\!B}^-,{\bf 0})$ and that $(P_{\!B}-k_B)^2 =
m^2$, this factor is
\begin{equation}
G\delta_{ij}\, {1-x_B \over {\bf k}_B^2 + m^2 - x_B(1-x_B)M^2}\,,
\end{equation}
where $x_B = k_B^-/P_{\!B}^-$ and ${\bf k}_B$ is the transverse part
of $k_B$. Comparing to the definition (\ref{psi}) of the meson wave
function $\psi$, this is
\begin{equation}
{1 \over x_B}\, \psi(x_B,{\bf k}_B^2)_{ij}\ .
\end{equation}
Thus when we square the amplitude and integrate over ${\bf k}_B$ we
obtain the parton distribution function for meson $B$,
Eq.~(\ref{pdf}).

In summary, the graph of Fig.~\ref{goodgraphfig} factors in a simple
way.  In the next section, we examine graphs that do not factor.

\section{Color exchange with spectator parton}

The graph in Fig.~\ref{spectatorfig} is similar to that in
Fig.~\ref{goodgraphfig}, but here the second gluon is absorbed on the
spectator antiquark line from meson B rather than emitted into the
final state.  As in Fig.~\ref{goodgraphfig}, there are eight ways of
attaching the gluons to the scalar quark and antiquark lines
associated with meson A; and there are four lowest order graphs for
the hard quark-gluon scattering. Thus one needs to sum over the 32
Feynman graphs contributing to the amplitude.

We will find two important differences between the cross section
associated with  Fig.~\ref{goodgraphfig} and that associated with
Fig.~\ref{spectatorfig}. First, there is an important observable
signature associated with Fig.~\ref{spectatorfig}.  Essentially all of
the plus-momentum $P_{\!A}^+ - P_{\!A}^{\prime +} \equiv z
P_{\!A}^+$ transferred from meson A appears in the system of high
$E_T$ jets. A negligible amount is transferred to the spectator
antiquark. Second, the cross section corresponding to graphs does not
take the factored form (\ref{factor}). The cross section is not simply
proportional to the probability to find the active partons from
mesons A and B, but also contains information on the spectator
interaction.

\begin{figure}[htb]
\centerline{\hskip 2cm \DESepsf(spectator.epsf width 8 cm) }
\smallskip
\caption{Graph with spectator parton interaction.}
\label{spectatorfig}
\end{figure}

Consider the integration over $q^\mu$ in Fig.~\ref{spectatorfig}.
Using power counting arguments similar to those in
Ref.~\cite{factorization}, one finds that the dominant integration
region for the transverse components of $q^\mu$ is ${\bf q}^2 \sim
m^2$.  For ${\bf q}^2$ much larger than this, the propagators in the
loop for meson A are thrown far off shell, while for ${\bf q}^2$ much
smaller than this there is a cancellation between attachments to the
quark and antiquark lines in meson A, since meson A is a color
singlet.  For $q^-$, the dominant integration region is $q^- \sim
m^2/P_{\!A}^+$, since, for $q^-$ much larger than this the
propagators in the loop for meson A are again thrown far off shell.
Similarly, for $q^+$, the dominant integration region is $q^+ \sim
m^2/P_{\!B}^-$ since, for $q^+$ much larger than this the
propagators for the quark and antiquark in meson B are thrown far off
shell. This integration region is denoted the Glauber region in
Ref.~\cite{factorization}.

We now examine more carefully the integration over $q^+$. In the
Glauber region, the only two propagators that have a significant
dependence on $q^+$ are the spectator-antiquark propagator
and the active-quark propagator.  These have the form
\begin{equation}
{i \over (P_{\!B}-k_B-q)^2 - m^2 + i \epsilon}
\approx
{i \over - 2(1 - x_B)P_{\!B}^- q^+
-\Lambda_1^2 + i \epsilon}
\end{equation}
and
\begin{equation}
{i \over (k_B + q)^2 - m^2 + i \epsilon}
\approx
{i \over 2x_B P_{\!B}^- q^+
- \Lambda_2^2 + i \epsilon}\,,
\end{equation}
where $x_B = {k_B^- / P_{\!B}^-}$ and
\begin{eqnarray}
\Lambda_1^2 &=&({\bf k}_B + {\bf q})^2  -{\bf k}_B^2 \,,
\nonumber\\
\Lambda_2^2 &=&
{x_B \over 1 - x_B}\ ({ \bf k}_B^2 + m^2)
+ ({\bf k}_B + {\bf q})^2 + m^2
 -x_B M^2\,.
\label{lambdadefs}
\end{eqnarray}
Notice that $\Lambda_1^2$ and $\Lambda_2^2$ are each of order $m^2$ in
the dominant integration region, in which $x_B$ is of order $1$ and
all ${\bf k}_B^2$ and ${\bf q}^2$ are of order $m^2$.

The crucial fact about these propagators is that the poles in $q^+$
are on opposite sides of the integration contour along the real $q^+$
axis, creating a pinch singularity as discussed in
\cite{factorization}. We can put the integral in a useful form by
deforming the contour into the upper half $q^+$ plane, past the pole
from the spectator propagator.  The deformation can be continued
until $q^+$ is large.  Then the propagators are far off shell, so
that the contribution from the deformed contour is negligible.  The
contribution that survives comes from the residue at the spectator
pole. Thus we make the replacement $q^+ \to -\Lambda_1^2 / [2 (1 -
x_B) P_{\!B}^-]$ in the active quark propagator.  This value of $q^+$ is
very small, and all other factors in the diagram depend only weakly
on $q^+$, so we may replace $q^+$ by $0$ everywhere else in the
diagram. Thus we make the replacements
\begin{eqnarray}
{i \over (k_B + q)^2 - m^2 + i \epsilon}
&\to&
-i \left[{x_B\over 1-x_B} \Lambda_1^2
+ \Lambda_2^2 \right]^{-1}
\nonumber\\
&=&
{-i\ (1-x_B)\over{({\bf k}_B + {\bf q})^2 + m^2}
- x_B(1-x_B) M^2 }
\end{eqnarray}
followed by
\begin{equation}
{i \over (P_{\!B}-k_B-q)^2 - m^2 + i \epsilon}
\to
{1 \over 2(1 - x_B)P_{\!B}^-}\
2\pi\, \delta\!\left(q^+\right).
\end{equation}
Finally, the interaction of the Glauber gluon with the spectator
quark is
\begin{equation}
- ig\,(2P_{\!B} - 2k_B -q)^\mu t_a\,,
\end{equation}
where $\mu$ is the polarization index of the gluon and
$a$ is its color index. This can be approximated by
\begin{equation}
- ig\, 2(1-x_B)P_{\!B}^- u^\mu t_a.
\end{equation}
Here $u^\mu = (0,1,{\bf 0})$ is a vector in the minus direction, so
$u\cdot A = A^+$.

What is the net result? The part of the diagram referring to
meson B and the interaction with the Glauber gluon begins as
\begin{eqnarray}
\lefteqn{F_a^\mu(k_B,q) =}
\\
&&\hskip - 0.7 cm
{i \over (k_B + q)^2 - m^2 + i \epsilon}
iG
{i \over (P_{\!B}-k_B-q)^2 - m^2 + i \epsilon}\,
(-ig)(2P_{\!B} - 2k_B -q)^\mu t_a\,.
\nonumber
\end{eqnarray}
With the replacements described above, this factor becomes
\begin{equation}
F_a^\mu(k_B,q) =
-{G\ (1-x_B)\over{({\bf k}_B + {\bf q})^2 + m^2}
- x_B(1-x_B) M^2 }\
2\pi\, \delta\!\left(q^+\right)\,
ig u^\mu t_a\,.
\end{equation}
We can write $F$ as
\begin{equation}
F_a^\mu(k_B,q) =
-{1 \over x_B}\,\psi(x_B,{\bf k}_B+{\bf q})\
{\cal J}_a^\mu(q)\,,
\label{Fform}
\end{equation}
where
\begin{equation}
{\cal J}_a^\mu(q) =
2\pi\, \delta\!\left(q^+\right)\,ig u^\mu t_a.
\end{equation}
and $\psi$ is the meson wave function, Eq.~(\ref{psi}).

We can understand this as follows. Let $A_a^\mu(q)$ represent the
lower half of the graph.  ($A$ depends on other variables, but we
suppress this dependence here and think of $A$ as a classical gluon
field.) Then the factor
\begin{equation}
\int {d^4q \over (2\pi)^4}
F^a_\mu(k_B,q)
A_a^\mu(q)
\label{topfactor}
\end{equation}
is approximated by
\begin{equation}
\int {d^4q \over (2\pi)^4}\
{-1 \over \, x_B}\, \psi(x_B,{\bf k}_B+{\bf q})\
2\pi\, \delta\!\left(q^+\right)\,ig u_\mu t_a\,
A_a^\mu(q)\,.
\label{kspace}
\end{equation}
Define Fourier transforms by
\begin{eqnarray}
A_a^\mu(q) &=&
\int d^4 x\ e^{iq\cdot x}\
\tilde A_a^\mu(x)\,,
\nonumber\\
\psi(x_B,{\bf k}_B+{\bf q})&=&
\int d^2 {\bf r}\ e^{-i ({\bf k}_B+{\bf q})\cdot{\bf r}}\
\tilde\psi(x_B,{\bf r})\,.
\end{eqnarray}
Then the factor in Eq.~(\ref{topfactor}) can be written as
\begin{equation}
-{1 \over x_B}\,
\int d^2{\bf r}\
e^{-i{\bf k}_B\cdot{\bf r}}\
\tilde\psi(x_B,{\bf r})\,
\int_{-\infty}^\infty dx^-
ig t_a\,
\tilde A_a^+(0,x^-,-{\bf r})\,.
\label{spectator}
\end{equation}
One can interpret Eq.~(\ref{spectator}) as follows.  The amplitude to
find a quark at the annihilation point (say $x^\nu = 0$) with
momentum fraction $x_B$ while the antiquark carries transverse
momentum $-{\bf k}_B$ is $\int d^2{\bf r}\ \exp({-i{\bf k}_B\cdot{\bf
r}})\, \tilde\psi(x_B,{\bf r})$. For the present graph, this amplitude
is modified by multiplication by a line integral of the color potential
along a line in the minus-direction at the transverse coordinate of
the antiquark, $-{\bf r}$ and the plus-coordinate of the antiquark
$x^+ \approx 0$. The antiquark is not appreciably deflected, but
provides a color current that can absorb the gluon that was needed to
balance the color and maintain meson $A$ as a color singlet. The
overall minus sign in Eq.~(\ref{spectator}) arises because the
spectator parton is an antiquark instead of a quark.

\begin{figure}[htb]
\centerline{ \DESepsf(spectatorsq.epsf width 10 cm) }
\smallskip
\caption{Cut graphs for inclusive jet production. Graph a)
contributes to diffractive jet production, while graph b) does not.}
\label{spectatorsqfig}
\end{figure}

We have seen that the contribution to the cross section calculated from
the graph of Fig.~\ref{spectatorfig} does not take the factorized form
of Eq.~(\ref{factor}). To see why the usual factorization  theorem
\cite{factorization} does not apply here, consider the square of
Fig.~\ref{spectatorfig}, which is represented by the cut Feynman diagram
of Fig.~\ref{spectatorsqfig}(a). If we were considering an inclusive jet
production cross section, then other graphs would also be present and
would cancel Fig.~\ref{spectatorsqfig}(a) to leading order in $m/E_T$.
These other graphs do not contribute to the cross section for
diffractive jet production. For instance, one graph that contributes to
inclusive jet production but not to diffractive jet production is shown
in Fig.~\ref{spectatorsqfig}(b). In this graph, the constituents of
hadron $A$, rather than a diffractively scattered version of hadron $A$,
appear in the final state.

\section{Color exchange with active parton}

We now consider graphs in which the gluon from meson $A$ is absorbed
on the active quark from meson $B$ before it enters the hard
interaction. A typical graph of this sort is shown in
Fig.~\ref{activegraphfig}. For convenience, we shall refer to the
gluon from meson $A$ as the Glauber gluon, although the important
integration region for its momentum $q^\mu$ is, strictly speaking,
not confined to the Glauber region as described in the previous
section. We will concentrate on the factor associated with the
absorption of the Glauber gluon on the active quark line from hadron
$B$,
\begin{equation}
{\cal G}_I \equiv
{i \over (k_B + q)^2 - m^2 + i \epsilon}\,
ig\,(2k_B + q)^\mu t_a\,.
\label{activestart}
\end{equation}
As in the case of Fig.~\ref{spectatorfig}, the dominant integration
region for the transverse components of $q^\mu$ is ${\bf q}^2 \sim
m^2$ and the dominant integration region for $q^-$ is $q^- \sim
m^2/P_{\!A}^+$.

\begin{figure}[htb]
\centerline{\hskip 2 cm \DESepsf(active.epsf width 8 cm) }
\smallskip
\caption{Graph with active parton interaction.}
\label{activegraphfig}
\end{figure}

The factor $ig\,(2k_B + q)^\mu t_a$ in (\ref{activestart}) multiplies a
vector associated with meson $A$ that is predominantly in the plus
direction.  Thus we can replace this factor by $ig\,2x_B P_{\!B}^-
u^\mu t_a$ with $x_B = {k_B^- / P_{\!B}^-}$ and $u^\mu = (0,1,{\bf
0})$. We now examine the integration over $q^+$. The only propagator
that has a significant dependence on $q^+$ for small $q^+$ is the active
quark propagator in (\ref{activestart}),
\begin{equation}
{i \over (k_B + q)^2 - m^2 + i \epsilon}
\approx
{i \over 2x_B P_{\!B}^- q^+
- \Lambda_2^2 + i \epsilon}\,,
\label{activeprop}
\end{equation}
where $\Lambda_2^2$ is given in Eq.~(\ref{lambdadefs}). Recall that
$\Lambda_2$ is of order $m^2$ in the dominant integration region. Thus
there is a pole in the lower half $q^+$ plane near $q^+ = 0$.
However, as discussed in Ref.~\cite{factorization}, the integration
contour is not pinched between two nearby poles. Thus the contour
can be deformed away from the pole. We then see that it is a good
approximation to replace $\Lambda_2^2$ by $0$ in (\ref{activeprop}).
Then
\begin{equation}
{\cal G}_I \approx
ig u^\mu t_a
{i \over q^+  + i \epsilon}\,.
\label{activemid}
\end{equation}

\begin{figure}[htb]
\centerline{ \DESepsf(activemod.epsf width 12 cm) }
\smallskip
\caption{Graphs with final state jet interaction.}
\label{activemodfig}
\end{figure}

Before completing work on the interaction with the active quark
entering the hard interaction, we need to consider the two
graphs, as shown in Fig.~\ref{activemodfig}, in which the gluon attaches
to one of the partons emerging from the hard scattering. Here the
factor associated with the absorption of the gluon is approximately
\begin{equation}
{\cal G}_F \approx
ig\,(2k_F - q)^\mu t_a
{i \over (k_F - q)^2 - m_F^2 + i \epsilon}\,,
\label{finalstart}
\end{equation}
where $t_a$ is the appropriate color matrix, depending on whether
the final state parton in question is a quark or a gluon, and where
$m_F$ is $m$ for a quark or $0$ for a gluon. This is
exactly the vertex factor for the graph in which the Glauber gluon is
absorbed on the outgoing quark line.  It is approximate in the case
that the gluon  with small momentum $q^\mu$ is absorbed on the
outgoing gluon line that carries large momentum $k_F$ into the final
state. As before, the factor $ ig\,(2k_F - q)^\mu t_a $ can be
approximated by $ ig\, 2k_F^- u^\mu t_a$. The propagator can be
approximated by
\begin{eqnarray}
{i \over (k_F - q)^2 - m_F^2 + i \epsilon}
&=&
{i \over - 2 k_F\cdot q + q^2 + i \epsilon}
\nonumber\\
&\approx& {i \over - 2 k_F\cdot q + i \epsilon}
\nonumber\\
&=&{i \over -2 k_F^- q^+ - 2 k_F^+ q^- + 2 {\bf k}_F\cdot {\bf q} +
i \epsilon}\,.
\end{eqnarray}
This propagator has a pole in the upper half $q^+$ plane. The poles
in the $q^+$ plane from the other propagators are all at large $q^+$,
$q^+ \sim E_T$. Thus we can deform the contour into the lower half
$q^+$ plane until $|q^+| \gg m$.  Now the components of $k_F$ are all
of the same order of magnitude (about $E_T$), while the minus and
transverse components of $q$ are both small.  Thus on the deformed
contour, the approximation
\begin{equation}
{i \over (k_F - q)^2 - m_F^2 + i \epsilon} \approx
{i \over  -2 k_F^-q^+ + i \epsilon}\,.
\label{finalprop}
\end{equation}
is valid.  Of course, the integral is independent of the contour
deformation, so we can move the contour back to the real $q^+$ axis.
(Similar manipulations may be found in Ref.~\cite{factorization}.)

With these approximations, ${\cal G}_F$ is
\begin{equation}
{\cal G}_F \approx
- ig u^\mu t_a
{i \over q^+  - i \epsilon}\,.
\label{finalmid}
\end{equation}

Notice that there is a certain similarity among the effective factor
${\cal G}_I$ giving the coupling of the Glauber gluon to the incoming
parton from hadron $B$ and the two factors ${\cal G}_F$ giving the
coupling to the outgoing partons. For the moment, let us consider
$q^+ \ne 0$, so that we can ignore the $i\epsilon$.  Then the only
difference is how the color matrix $t_a$ for the Glauber gluon is
connected to the color matrix $M^{i\, j}_{c\,d}$ for the hard
scattering. Here the color indices are $i$ for the final quark, $j$
is the initial quark, $c$ for the final gluon, and $d$ for the
initial gluon. Attaching to the final quark, we have
$(t^{(3)}_a)_{ii^\prime}\, M^{i^\prime j}_{c\,d}$, where the
superscript $(3)$ denotes the triplet representation of SU(3).
Attaching to the final gluon, we have $(t^{(8)}_a)_{cc^\prime}\,
M^{i\, j}_{c^\prime d}$.  Attaching to the initial quark, we have
$- M^{i\, j^\prime}_{c\,d}\,(t^{(3)}_a)_{j^\prime j}$, where we have
noted the opposite signs between Eq.~(\ref{activemid}) and
Eq.~(\ref{finalmid}).  We can sum the three contributions using the
group theoretic identity
\begin{equation}
(t^{(3)}_a)_{ii^\prime}\, M^{i^\prime j}_{c\,d}
+(t^{(8)}_a)_{cc^\prime}\, M^{i\, j}_{c^\prime d}
- M^{i\, j^\prime}_{c\,d}\,(t^{(3)}_a)_{j^\prime j}
= M^{i\, j}_{c\,d^\prime}\,(t^{(8)}_a)_{d^\prime d}\,,
\label{colorident}
\end{equation}
which merely says that $M$ is invariant under SU(3) rotations.  In
the color matrix on the right hand side of (\ref{colorident}), the
Glauber gluon from hadron $A$ first couples to the active gluon from
hadron $A$ to make a color octet object, then this color octet couples
to the hard scattering color matrix.  However, these two gluons are
in a color singlet state, so coupling the two to make a color octet
gives zero.  (That is, the lower part of the graph is proportional to
$\delta_{ad}$, while $(t^{(8)}_a)_{d^\prime d}$ is antisymmetric in
$\{a,d\}$.)

Thus the sum of the three diagrams cancels, except for the fact that
the denominators in Eqs.~(\ref{activemid}) and
(\ref{finalmid}) have the opposite $i\epsilon$'s.
In order to take the $i\epsilon$'s into account we now make the
replacement
\begin{equation}
{i \over q^+  + i \epsilon} \to
{i \over q^+  - i \epsilon}
+ 2 \pi \delta(q^+)\,.
\label{mysterious}
\end{equation}
in ${\cal G}_I$. The first term has the right $i\epsilon$ to
participate in the cancellation, and we are left with an uncancelled
term
\begin{equation}
{\cal G}_I^\prime \equiv
ig u^\mu t_a
2 \pi \delta(q^+)\,.
\label{activefinal}
\end{equation}
We are grateful to John Collins for suggesting this to us.

What is the net result? The part of the diagram referring to
meson B and the interaction with the Glauber gluon begins as
\begin{equation}
F_a^{\prime\mu}(k_B,q) =
{i \over (k_B+q)^2 - m^2 + i \epsilon}\
ig\,(2k_B +q)^\mu t_a
{i \over k_B^2 - m^2 + i \epsilon}
iG\,.
\nonumber
\end{equation}
With the replacements described above, this factor becomes
\begin{equation}
F_a^{\prime\mu}(k_B,q) =
{G\ (1-x_B)\over ({\bf k}_B + m^2)
- x_B(1-x_B) M^2 }\
2\pi\, \delta\!\left(q^+\right)\,
ig u^\mu t_a\,.
\end{equation}
We can write $F^\prime$ as
\begin{equation}
F_a^{\prime\mu}(k_B,q) =
{1 \over x_B}\,\psi(x_B,{\bf k}_B)\
{\cal J}_a^\mu(q)\,,
\end{equation}
where ${\cal J}$ and $\psi$ are defined as in Eq.~(\ref{Fform}).

It is instructive to write this in transverse coordinate space, as
we did for the attachment to the spectator antiquark. We let
$A_a^\mu(q)$ represent the lower half of the graph. Then the factor
\begin{equation}
\int {d^4q \over (2\pi)^4}
F_a^{\prime \mu}(k_B,q)
A^a_\mu(q)
\label{topfactorbis}
\end{equation}
is approximated by
\begin{equation}
\int {d^4q \over (2\pi)^4}\
{1 \over x_B}\, \psi(x_B,{\bf k}_B)\
2\pi\, \delta\!\left(q^+\right)\,ig u_\mu t_a\,
A_a^\mu(q)\,.
\label{kspacebis}
\end{equation}
Define Fourier transforms as before, we find that the factor in
Eq.~(\ref{topfactorbis}) becomes
\begin{equation}
{1 \over x_B}\,
\int d^2{\bf r}\
e^{-i{\bf k}_B\cdot{\bf r}}\
\tilde\psi(x_B,{\bf r})\,
\int_{-\infty}^\infty dx^-
ig t_a\,
\tilde A_a^+(0,x^-,{\bf 0})\,.
\label{active}
\end{equation}
This is the same as the factor (\ref{spectator}), except for the
overall sign and except that the Glauber gluon is absorbed at
transverse position $\bf 0$ instead of transverse position $-{\bf r}$.

\section{Results}

We have investigated diffractive jet production within the context of a
certain simple model. We have seen that the cross section has
contributions that are lossless in the sense that the plus-momentum
$zP_{\!A}^+$ transferred from hadron $A$ is transferred without loss
to the jet system. Thus these terms contain a factor $\delta(1-X_A/z)$.
We now assemble the results and rewrite them in a suggestive form.

We can build up the result using four ingredients.  The first
ingredient is the line integral of the color potential along a
lightlike line in the minus direction (the direction of hadron $B$)
at a transverse position $\bf b$, where the hard interaction is at
transverse position $\bf 0$:
\begin{equation}
{\cal A}({\bf b}) =
\int_{-\infty}^{+\infty}\!\! d y^-
\sum_c ig t_c A_c^+(0,y^-,{\bf b})\,.
\end{equation}
Notice that, in addition to being an operator, ${\cal A}$ is a
matrix that acts on vectors in the $\bf 3$-representation of color
SU(3). The next ingredient is the color field operator
$F_a^{+j}(0,0,{\bf 0})$, which destroys a gluon of transverse
polarization $j$ and color $a$ at the origin of
space-time. Combining these operators, we form the amplitude to
annihilate the gluon at the origin along with one more Glauber gluon
at transverse position $\bf b$, while scattering the meson $A$ with
momentum transfer $(z,t)$:
\begin{equation}
G^j_a({\bf b};t,z) =
{1 \over 4\pi X_A P_{\!A}^+}\,
\langle P_{\!A}^\prime |
T\left\{
{\cal A}({\bf b})\,
F_a^{+j}(0,0,{\bf 0})
\right\}
|P_{\!A} \rangle\,.
\end{equation}
(We have indicated a time-ordered product here, but within the
context of the model the operators commute, so the time-ordering is
not relevant.)  The next ingredient is the wave function to find the
quark in meson $B$ carrying a momentum fraction $X_B$ and separated
from the antiquark by a transverse separation $\bf r$, $\psi(X_B,{\bf
r})$.

Finally, we have a hard scattering function
$H^{jk}_{ab}\left(\hat s,E_T\right)_{IJ}$, where $\hat s = X_A X_B
s$.  Up to a normalization, this function is the product of the
Born-level scattering amplitude $\cal M$ for the gluon of
plus-momentum $x_A P_{\!A}^+$ to scatter from a (scalar) quark of
minus-momentum $x_B P_{\!B}^-$ times the complex conjugate amplitude
${\cal M}^\dagger$ with different color and spin labels. The gluon has
of polarization $k$ and color $b$, while the quark has color $J$.  In
the complex conjugate amplitude, the corresponding indices are $j$,
$a$, and $I$. We consider $H$ as a matrix in the quark color space
and do not write the indices $IJ$ explicitly. The normalization is
such that the usual hard scattering cross section $d \hat \sigma$ for
inclusive jet production, as used in Eq.~(\ref{usualfact}), is, at
the Born level, the color and spin average of $H$, with the
momentum-conserving delta functions removed:
\begin{equation}
\left[
{d \hat\sigma(a + b\to {\rm jets} +X) \over d E_T\, dX_A\, dX_B}
\right]_{\rm Born}
 =
\delta(x_A-X_A)\,\delta(x_B-X_B)\,
{1 \over 2}\sum_j
{1 \over 8}\sum_a
{1 \over 3}{\rm Tr}\left\{H_{aa}^{jj} \right\}.
\end{equation}

With these ingredients, the lossless part of the cross
section for diffractive jet production within the model described in
the preceding sections is
\pagebreak[2]
\begin{eqnarray}
\left[{d \sigma^{\rm diff}(A + B\to A +{\rm jets} +X)\over
d E_T\, dX_A\, dX_B\, dz\, dt}\right]_{0}
&\sim&
\delta\left( 1-X_A/z  \right)
\int d{\bf r}\,
{|\psi(X_B,{\bf r})|^2 \over 2X_B(1-X_B)}
\nonumber\\ &&\hskip -4cm \times
\sum_{j,k = 1}^2 \sum_{a,b = 1}^8 {\rm Tr} \biggl\{
\left[ G^j_a(-{\bf r};t,z) - G^j_a({\bf 0};t,z)
\right]^\dagger H^{jk}_{ab}\left(\hat s,E_T\right)
\nonumber\\ && \times
\left[ G^k_b(-{\bf r};t,z) - G^k_b({\bf 0};t,z)
\right]\biggr\}.
\label{result}
\end{eqnarray}

There are several features of this result that are notable.  First is
the $\delta(1-X_A/z)$, which provides the experimental signature for
the lossless contribution to the cross section.  In contrast, the
normal contributions yield values of the variable $X_A$ that are
spread over the region $X_A < z$.  The second feature is that the
factorization that applies to inclusive jet production does not apply
here. The factor corresponding to hadron $A$ is tied to the factor
representing hadron $B$ by a convolution over a transverse position
$\bf r$.  There is also a more complicated color and spin structure
than in the normal contributions.

We emphasize that the fact that factorization does not apply here does
not represent a failure of the factorization theorem for the total
cross section for $A + B \to {\rm jets} + X$. The factorization
theorem requires an unrestricted sum over the final state $X$
\cite{factorization}.  When we demand that $X$ contain the a
diffractively scattered hadron $A$, the conditions for the theorem are
violated.

Despite its rather complicated structure, the interpretation of
Eq.~(\ref{result}) is straightforward. In the model, meson $B$
consists of a quark and an antiquark.  With probability $\propto
|\psi(X_B,{\bf r})|^2$, they are separated by a transverse distance
$\bf r$. In order to restore the color of hadron $A$, we must absorb a
gluon on either the antiquark (at position $-{\bf r}$) or the quark
(at position $\bf 0$). Since the quark and antiquark have opposite
color charges, the absorption amplitude is proportional to the
difference $G^j_a(-{\bf r};t,z) - G^j_a({\bf 0};t,z)$. Here
$G^j_a({\bf b};t,z)$ is the amplitude to absorb a color field
quantum at transverse position ${\bf b}$ when the ``active'' gluon is
annihilated at the origin of space-time and hadron $A$ is
diffractively scattered.  Thus $G$ describes the color field
associated with the pomeron when one gluon from the pomeron has been
annihilated at the origin.

Here we meet an interesting experimental possibility.  The ${\bf b}$
dependence of $G^j_a({\bf b};t,z)$ reflects the transverse structure
of the pomeron. It has significant structure on some distance scale
$R_{\rm P}$ characteristic of the pomeron. In the present
model, $1/R_{\rm P}$ is of order of the quark mass $m$. Thus
$G^j_a(-{\bf r};t,z) - G^j_a({\bf 0};t,z)$ is small when $|{\bf r}|
\ll R_{\rm P}$.  On the other hand, $|\psi(X_B,{\bf r})|^2$ is small
when $|{\bf r}| \gg R_B$, where $R_B$ is a characteristic size of
hadron $B$.  This size is also of order $1/m$ in the model. However,
suppose that we generalize the model so that $R_B$ can be separately
adjusted.  Then when $R_B \sim  R_{\rm P}$, there will be a
substantial contribution to the cross section proportional to
$\delta\left( 1-X_A/z \right)$.  But when $R_B \ll  R_{\rm P}$, this
contribution will vanish.

So far, we have worked only with a simple model.  But the model
suggests a plausible conjecture.  First, there can be a sizable
contribution to diffractive jet production proportional to
$\delta\left( 1-X_A/z \right)$, arising from using one gluon from the
pomeron to make the jets and absorbing  on the partons of hadron $B$
the rest of the color field needed to make hadron $A$ back into a
color singlet. Second, when the size $R_B$ of hadron $B$ is small
compared to the transverse size $R_{\rm P}$ associated with the color
field in pomeron exchange, then hadron $B$ should act as a color
singlet and this contribution should disappear.

In order to test this conjecture, and probe the transverse structure of
the pomeron, one needs to use hadrons of adjustable size.  This is
simple (for theorists).  At HERA, one manufactures bremsstrahlung
photons from the electron beam.  The virtuality  $Q = [{-P_{\!B}^\mu
P_{\!B\mu}}]^{1/2}$ of the photon is measured by the deflection of the
electron, and can be anything from nearly zero to many GeV. The photon
can collide with a proton (hadron $A$) to make jets with
$E_T\gg Q$. The cross section for this process can be (roughly) divided
into two parts. In one part, the photon acts as a parton and scatters
directly with a parton from hadron $A$ to make the jets.  In the other
part, the photon acts as a hadron, made of constituent partons. For $Q
\approx 0$, this hadron is essentially a $\rho$-meson, with a size $R_B
\approx 1 {\ \rm fm}$.  For $Q \gg 1 {\ \rm fm}^{-1}$, the ``hadron''
consists of a quark-antiquark pair, with wave functions given in
Eq.~(\ref{psiT}) for transverse polarization and Eq.~(\ref{psiL}) for
longitudinal polarization. These wave functions are characterized by a
size $R_B \approx [X_B (1-X_B) Q^2]^{-1/2}$. Since $Q^2$ and
$X_B$ are measurable, this size is adjustable.

Thus the experimental possibility is to select photon-proton
scattering events with $Q$ in the range from zero up to one or two GeV
and look for jets accompanied by a diffractively scattered proton.
These can be divided into events with a partonic photon ($X_B \approx
1$) and those with a hadronic photon ($X_B<1$). Some of the events
should be of the lossless ($X_A \approx z$) type, while some most should
be of the normal ($X_A <z$) type . For the partonic-photon events, there
should be diffractive jet production, but it should have $X_A < z$,
since the photon has no color structure. For the hadronic-photon
events, both $X_A <z$ and $X_A \approx z$ events should occur, but the
fraction of $X_A \approx z$ events should approach zero as $Q$ becomes
large. The scale of $X_B (1-X_B) Q^2$ at which this happens  reflects
the transverse size of the pomeron.

In the model we have used, the ``lossless'' contributions to the cross
section are proportional to $\delta(1-X_A/z)$ while the ``normal''
contributions are non-singular functions of $X_A/z$. We expect that
this clean division is not so clean when higher-order contributions are
taken into account.  Presumably there are contributions that are
singular as $X_A/z \to 1$ but are not as singular as a delta-function.
Thus one may anticipate that the lowest order $\delta(1-X_A/z)$
appears as events with  $X_A \approx z$ rather than $X_A = z$.

We must emphasize that the proposal given above is a conjecture based
on a simple model, not a proven consequence of QCD.  It should be a
challenge to investigate the structure of diffractive hard scattering
further and to discover what features of the model survive a higher
order analysis.
\acknowledgements
We thank J.\ C.\ Collins and J.\ F.\ Gunion for helpful conversations.
This work was supported by the U.S.\ Department of Energy.


\end{document}